\documentclass[prb, reprint, superscriptaddress]{revtex4-1}
\usepackage{graphicx}
\usepackage{amsmath}
\usepackage{amssymb}

\usepackage[T1]{fontenc}
\usepackage{xcolor}
\usepackage[colorlinks, 
            linkcolor={blue},
            citecolor={blue},
            urlcolor={blue}
            ]{hyperref}
\usepackage[normalem]{ulem}

\makeatletter
\newsavebox{\@brx} 
\newcommand{\llangle}[1][]{\savebox{\@brx}{\(\m@th{#1\langle}\)}%
  \mathopen{\copy\@brx\mkern2mu\kern-0.9\wd\@brx\usebox{\@brx}}}
\newcommand{\rrangle}[1][]{\savebox{\@brx}{\(\m@th{#1\rangle}\)}%
  \mathclose{\copy\@brx\mkern2mu\kern-0.9\wd\@brx\usebox{\@brx}}}
\makeatother

\begin{document}

\title{%
Theoretical description of time-resolved photoemission in
charge-density-wave materials out to long times
}

\author{Marko D. Petrovi{\'c}}
\email{mp1770{@}georgetown.edu} 
\affiliation{Department of Physics, Georgetown University,
               Washington, DC 20057, USA}
\author{Manuel Weber}
\affiliation{Max Planck Institute for the Physics of
             Complex Systems, N{\"o}thnitzer Str. 38,
             01187 Dresden, Germany}
\author{James K. Freericks}
\email{James.Freericks{@}georgetown.edu}
\affiliation{Department of Physics, Georgetown University,
               Washington, DC 20057, USA}

\date{\today}

\begin{abstract}
  We describe coupled electron-phonon systems
  semiclassically---Ehrenfest dynamics for the phonons and
  quantum mechanics for the electrons---using a classical Monte
  Carlo approach that determines the nonequilibrium response to
  a large pump field. The semiclassical approach is quite
  accurate, because the phonons are excited to average energies
  much higher than the phonon frequency, eliminating the need
  for a quantum description. The numerical efficiency of this
  method allows us to perform a self-consistent time evolution
  out to very long times (tens of picoseconds) enabling us to
  model pump-probe experiments of a charge density wave (CDW)
  material. Our system is a half-filled, one-dimensional (1D)
  Holstein chain that exhibits CDW ordering due to a Peierls
  transition. The chain is subjected to a time-dependent
  electromagnetic pump field that excites it out of equilibrium,
  and then a second probe pulse is applied after a time delay.
  By evolving the system to long times, we capture the complete
  process of lattice excitation and subsequent relaxation to a
  new equilibrium, due to an exchange of energy between the
  electrons and the lattice, leading to lattice relaxation at
  finite temperatures. We employ an indirect (impulsive) driving
  mechanism of the lattice by the pump pulse due to the driving
  of the electrons by the pump field. We identify two driving
  regimes, where the pump can either cause small perturbations
  or completely invert the initial CDW order. Our work
  successfully describes the ringing of the amplitude mode in
  CDW systems that has long been seen in experiment, but never
  successfully explained by microscopic theory. 
\end{abstract}

\maketitle

Time-resolved angle-resolved photoemission
spectroscopy (trARPES) is an ultrafast measuring
technique~\cite{arpes_rev_2021} capable of capturing the
real-time evolution and occupation of electronic states in an
array of materials: from
superconductors,~\cite{gerber_science_2017} semimetals such as
graphene~\cite{aeschlimann_nlet_21} to topological
insulators.~\cite{sobota_prl_2012} A particular focus is given
to CDW materials~\cite{gruner_book, gruner_rev} due to the
specific ordering nature of their ground state and different
competing phases they can exhibit. The trARPES experiments
utilize a combination of time-delayed pump and probe pulses to
excite the studied material and track the dynamics of the
induced nonequilibrium state. The technique was so far
successfully used to study oscillations of the CDW amplitude
mode, as well as CDW melting in various materials, from $\rm
TaS_2$~\cite{perfetti_prl2006, perfetti_njp2008} to rare-earth
tritellurides such as $\rm TbTe_3$,~\cite{schmitt_sci2008,
schmitt_njp2011, maklar_ncomm2021} $\rm
LaTe_3$,~\cite{kogar_natphys_2020} and 
$\rm DyTe_3$.~\cite{zong_prl_2021}

In a typical trARPES experiment, the pump-induced modification
of the electronic structure is observed in the measured
photoemission spectrum (PES). The lattice can be driven due to
direct dipole coupling with the pump field or
indirectly through coupling with electrons. The conventional
explanation of the indirect driving involves the transfer of
energy from the laser pump to electrons, happening on a
femtosecond timescale, and then from electrons to the crystal
lattice, happening on a picosecond timescale. A detailed
understanding of this driving mechanism is essential in the
context of Floquet engineering~\cite{hubner_nlet_2018} where the
dressing of the electronic structure by lattice motion can be
used to fine-tune the electronic spectrum.

The theoretical description of pump-probe experiments for CDW
systems has relied on phenomenological approaches like
time-dependent Ginzburg-Landau theory, whereas the study of
microscopic models has been hindered by the absence of efficient
numerical methods that can resolve the very different timescales
of electron and phonon dynamics. Dynamical mean-field
theory~\cite{moritz_prb2010, matveev_prb2016} and related
studies~\cite{shen_prb2014, freericks_prl2014,
freericks_physc2017}
focused on the excitation of the nonequilibrium electronic state
for several tens of femtoseconds, but the slow lattice motion
was not considered. Exact diagonalization and the
density-matrix renormalization group can provide
exact results for small lattice sizes, short times scales, and
high phonon frequencies;~\cite{Filippis2012, matsueda,
hashimoto_prb, stolpp_prb, Sous2021} as time proceeds,
the growing number of phonon excitations renders simulations
impossible, in particular for phonon frequencies in the meV
range which are required for an accurate description
of most CDW materials. To explain the indirect driving
mechanism, as exemplified by the experiment in
Ref.~\onlinecite{schmitt_sci2008}, it is necessary 
to develop approaches that naturally include the slow phonon
dynamics. One possibility to achieve this is by the so-called
exact factorization,~\cite{gross_01}
while another is by the multitrajectory Ehrenfest
method.~\cite{lively_2021}

%
\begin{figure}[tb]
\begin{center}
\includegraphics[width=0.49\textwidth]{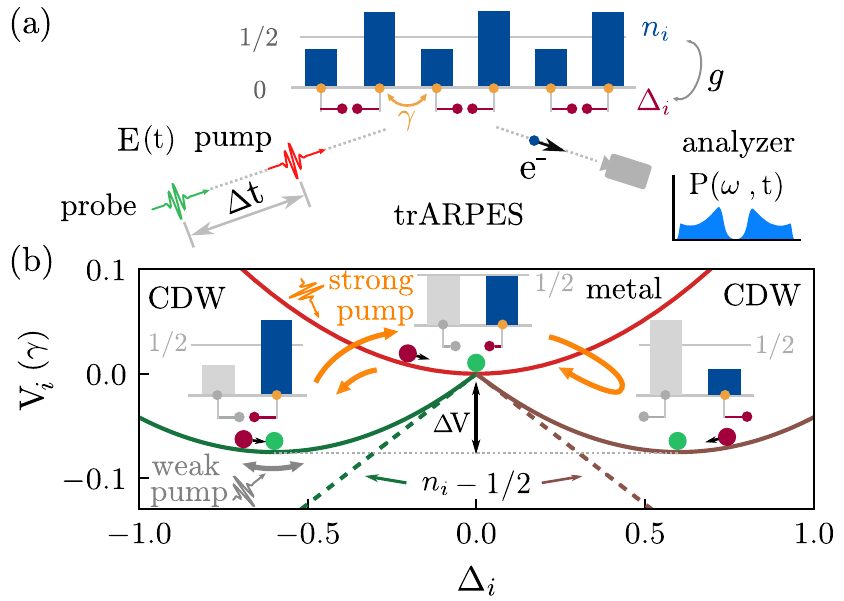}
\end{center}
\caption{\label{fig:system}
  (a) Schematics of the 1D Holstein chain of tight-binding sites
  (orange dots) with local lattice displacements $\Delta_i$ at
  every site (dark red dots and dark red lines below the orange
  dots) in our trARPES setup. A pump pulse $E(t)$ excites the
  electronic charge density $n_i$ (dark blue bars) by
  modifying the Peierls phase factor of the tight-binding
  hopping parameter $\gamma$. The energy distribution of the
  injected electrons is captured in the computed photoemission
  spectrum $P(\omega, t)$. (b) Three potential energy profiles
  (green parabola, red parabola, and brown parabola) of a single
  lattice displacement $\Delta_i$ (dark red circles and dark red
  lines on the right side of the three insets above the
  parabolas) for different values of the onsite charge density
  $n_i$ (blue bars on the right side of the three insets).
  Green dots show the potential minima of the three parabolas.
  The system is initialized at $T = 0$ temperature. The green
  and brown parabola with green dots also show the two possible
  distorted charge density wave states with negative and
  positive displacements, respectively. For the context of what
  happens on the neighboring sublattice, see the grayed sites
  and bars on the left side of the three insets. The two
  possible distorted states are separated by an energy barrier
  $\Delta V$. Pump excitation modifies the onsite charge density
  $n_i$, which rotates the potential energy parabola around the
  $\Delta_i = 0$ point (see the orange and gray arrows which
  depict the motion of potential minima for strong and weak pump
  excitation, respectively). This happens because the derivative
  of the potential (\ref{eq:pot}) at $\Delta_i = 0$ is $n_i -
  1/2$ (the dashed green and brown lines). For uniform charge
  density the system is considered to be metallic (inset above
  the red parabola) and the derivative is zero.
}
\end{figure}
%

In this article, we extend a time-dependent Monte Carlo
(MC) method,~\cite{weber_2021} recently developed for frozen
phonons, to include the classical lattice motion.
This approach is similar to that of ab-initio molecular
dynamics,~\cite{cp_method} except we focus more on the
measurable electronic properties in the context of trARPES
experiments. A classical description of the phonons
is expected to become accurate if temperature and/or the Peierls
gap are larger than the phonon frequency,~\cite{Brazovskii1976}
as confirmed by exact quantum Monte Carlo simulations in
equilibrium.~\cite{weber_prb2018} For realistic parameters,
this is already the case at very low temperatures.
Although the semiclassical approach is poor at
zero temperature, because it neglects the quantum fluctuations
in the ground state and has no damping, we find that a sliding
time average can explain some of the features we observe at
finite temperatures. Furthermore, the addition of energy into
the system via a strong pump only improves the accuracy of this
method, as the quantization of the phonon energies becomes
less and less important. For our method to be efficient, we
neglect electron-electron interactions; as they mainly affect
the short-time decay after the pump, we expect our method to
capture the universal long-time features of the induced
lattice dynamics. In particular, our method reveals all the
details of the indirect driving mechanism found in experiments.

Before explaining the
method, we briefly describe our system and summarize our main
findings. A simplified schematics of a trARPES setup to which we
apply our method is shown in Fig.~\ref{fig:system}(a). 
In equilibrium at zero temperature, it is a
perfectly-dimerized tight-binding chain in the CDW phase, where
alternating lattice displacements $\pm\Delta^{\rm eq}_i$ are
accompanied by an alternating charge density $n^{\rm eq}_i$
varying around 1/2. The indirect driving mechanism can be
explained by looking at the potential energy profile of
each lattice displacement, as illustrated in
Fig.~\ref{fig:system}(b). The pump perturbs the equilibrium
electronic order, thus changing the local
charge density $n_i(t)$. The perturbed electronic state sets a
new potential profile for the phonons at each site by rotating
the equilibrium potential parabola around $\Delta_i = 0$ and
setting a new dynamical minimum $\Delta^{\rm min}_{i}(t)$ (green
dots in Fig.~\ref{fig:system}(b)) towards which each local
displacement $\Delta_i(t)$ tends to move.

The pump can be considered strong or weak depending on its
amplitude and frequency. Strong pumps will drive the system from
the initial CDW state on the left inset of
Fig.~\ref{fig:system}(b) to an ``overshot'' state on the right
inset with an inverted order parameter. The system left to
itself after the pump will oscillate between these two
insulating CDW orders by briefly going through a metallic state
(the top inset in Fig.~\ref{fig:system}(b)). Flipping the CDW
order, as exemplified by the site considered in
Fig.~\ref{fig:system}(b), means the local charge density $n_i(t)$
goes below 1/2 (if initially being above 1/2), while the local
displacement $\Delta_i(t)$ changes sign. By contrast, weak
pumps only slightly perturb the initial {CDW}. They cause weak
oscillations of the local displacements $\Delta_i(t)$, but never
change their sign. We show both these scenarios, for the weak
and the strong pump, in two videos in the
Supporting Information Appendix (SI).

The pump-induced lattice motion will change the electronic
structure of the system, causing the insulating gap energy to
oscillate. The coupled dynamics will repeat until the system
reaches a new equilibrium.
This behaviour can be deduced from the computed PES
(Fig.~\ref{fig:PES}) as well as from the order parameters for
the electronic and lattice subsystems (Fig.~\ref{fig:order}).
Additionally, we explore what are the conditions in terms of
pump amplitude and frequency which would cause weak or strong
driving (Fig.~\ref{fig:melt}). In the pumping regime that we
consider, where the photon frequencies are at least one order of
magnitude higher than the phonon ones, pump driving is modulated
mostly by its amplitude. We find there is a threshold amplitude
at which the driving begins, and also another one which
determines the transition from the weak to the strong pumping
regime. We contrast this with a transient CDW melting regime
where the two subsystems are temporarily decoupled and the
lattice oscillates at its intrinsic frequency. Additionally, we
provide the details of electron dynamics and populations of the
two subbands during the pump pulse in the SI.

\section*{Overview of the Method}

The system illustrated in Fig.~\ref{fig:system}(a) is modeled 
by the 1D spinless Holstein model
  \begin{equation}
    \hat{H} = \hat{H}_{\rm el} + \hat{H}_{\rm ph} \,.
     \label{eq:ham}
  \end{equation}
The electronic subsystem consists of a nearest-neighbor
tight-binding model with uniform hopping $\gamma$ and an
electron-phonon interaction with coupling constant $g$, i.e.,
\begin{eqnarray}
   \hat{H}_{\rm el} =
   -\sum_{i=1}^{L}
   \left(\gamma \, \hat{c}^\dagger_i\hat{c}_{i+1}
        + {\textrm{h.c.}} \right) 
  +\ g \sum_{i=1}^{L}\hat{q}_{i}\left(
     \hat{n}_i - \frac{1}{2}\right).
     \label{eq:ele}
\end{eqnarray}
Here, $\hat{c}^\dagger_{i}$ and $\hat{c}_{i}$ are the electronic
creation and annihilation operators at lattice site $i$, while
$\hat{q}_i$ is the local phonon displacement operator coupled
to the local charge density $\hat{n}_{i} =
\hat{c}^\dagger_{i}\hat{c}_i$. We impose periodic boundary
conditions to eliminate edge effects. A time-varying pump field
is implemented via the Peierls substitution $\gamma \to \gamma
\, e^{-i\varphi(t)}$ on all sites, thus rendering the electronic
Hamiltonian $\hat{H}_{\rm el}(t)$ explicitly time-dependent.
Following Refs.~\onlinecite{freericks_pes,freericks_prl2014}, the
probe is considered only perturbatively and is not included in
the Peierls phase.
The second term in the total Hamiltonian consists of the phonon
kinetic and potential energy, 
\begin{equation}
   \hat{H}_{\rm ph} =
   \sum_{i=1}^{L} \frac{\hat{p}_{i}^{2}}{2M} +
   \sum_{i=1}^{L} \frac{K}{2}\hat{q}_{i}^{2} \, ,
   \label{eq:phe}
\end{equation}
with $K$ being the phonon spring constant and $M$ the phonon
mass; $\hat{p}_i$ is the phonon momentum operator. 
We also define the phonon frequency $\Omega = \sqrt{K / M}$
and the dimensionless electron-phonon coupling constant $\lambda
= g^2/(4K\gamma)$. Through the rest of our paper we use
$\gamma$ as the unit of energy and set $\Omega = 0.01 \, \gamma
/ \hbar$, $\lambda = 0.6$, and $L = 30$ if not stated otherwise.

To solve for the long-time dynamics of the coupled
electron-phonon system, we treat the electrons quantum
mechanically but approximate the phonons by classical variables
$q_i$ and $p_i$. A classical description of the phonons becomes
exact in the {\it frozen phonon}\ limit where $M \to \infty$
(i.e., $\Omega \to 0$); furthermore, we describe why the
classical description should also be quite accurate when the
average energy in the phonons is larger than $\hbar\Omega$ in
the {SI}. While the short-time response of the electrons
to an applied pump field can be calculated efficiently in the
static limit, the slow energy exchange between electrons and
phonons requires small but finite phonon frequencies. To this
end, we apply the Ehrenfest theorem to obtain the classical
phonon equations of motion \begin{eqnarray}
    \frac{d}{dt}{\Delta}_{i}(t) & = & \Omega^2 \pi_i(t) \, ,
     \label{eq:momenta_scaled}
       \\
     \frac{d}{dt}\pi_i(t) & = & -{\Delta}_{i}(t) -
      4\lambda\gamma\left(n_i(t) - 1/2 \right) \, ,
    \label{eq:force_scaled}
\end{eqnarray}
for the rescaled classical variables
$\Delta_i(t) = g \, q_i(t)$ and $\pi_i(t) = g \, p_i(t)/K$.
Here, $n_i(t) = \llangle \hat{n}_i(t)
\rrangle_{\vec{\Delta}_0,\vec{\pi}_0}$ is the electronic
expectation value of the charge density at time $t$, considering
that the phonons were initialized in a configuration
$(\vec{\Delta}_0, \vec{\pi}_0)$. The r.h.s.~of
Eq.~(\ref{eq:force_scaled}) defines a force that depends on the
local phonon displacement $\Delta_i(t)$ and on the charge
density $n_i(t)$.

For classical phonons, we calculate time-dependent observables
$\langle \hat{O}(t) \rangle
 = \int d\vec{\Delta}_0 \int d\vec{\pi}_0 \,
 W_\mathrm{eq}(\vec{\Delta}_0,\vec{\pi}_0) \,
 \llangle \hat{O}(t)
 \rrangle_{\vec{\Delta}_0,\vec{\pi}_0}
$
as a weighted average with respect to the equilibrium phonon
distribution $W_\mathrm{eq}$. We initialize our system in the
equilibrium solution of the static phonon limit, which is
accurate for temperatures $k_\mathrm{B} T \gtrsim \hbar
\Omega$.~\cite{weber_prb2018} Then, the
initial phonon displacements $\vec{\Delta}_0$ are sampled from
$W_\mathrm{eq}[\Omega=0]$ using a classical MC method
\cite{weber_prb2016} and we set $\vec{\pi}_0=0$. At $T=0$, our
system is set up by a single configuration $\Delta_{0,i}=(-1)^i
\Delta$ with perfect dimerization which is accompanied by CDW
order, as illustrated in Fig.~\ref{fig:system}(a). A band gap of
$2\Delta$ separates the fully-filled lower band from the empty
upper band, so that the rescaled displacements are directly
related to the single-particle gap $\Delta$. Although there is
no true long-range CDW order in 1D at $T>0$, the band gap as
well as short-range CDW correlations remain stable up to
$k_\mathrm{B} T \approx 0.1 \, \gamma$. For details on the
equilibrium solution, see Ref.~\onlinecite{weber_prb2016}.

For each initial phonon configuration, the coupled
electron-phonon dynamics is implemented self-consistently in two
steps. In the first step, we update the electron annihilation
operators $\hat{c}_i(t+\Delta t) =
\sum_{j=1}^{L}\mathcal{U}_{ij}(t+\Delta t, t)\hat{c}_j(t)$
through evolution by direct diagonalization (for technical
details see Ref.~\onlinecite{weber_2021}). This allows us to
update $n_i(t)$. In the second step, the obtained $n_i(t)$ is
replaced in Eq.~(\ref{eq:force_scaled}) and new lattice
displacements are computed using the Verlet integration scheme.
Because $\hat{H}_\mathrm{el}(t)$ is quadratic for each phonon
configuration, we only need $\mathcal{O}(L^3)$ operations for
every time step. This allows us to reach long enough times to
observe the damped phonon oscillations also found in
experiments. Note that the MC average over many initial phonon
configurations recovers the interacting nature of our
electron-phonon coupled system.

The chain is driven out of equilibrium with an external electric
field applied uniformly in space along the $x$ direction,
  \begin{equation}
    \mathbf{E}(\mathbf{r}, t) =
        E_{0}\exp{%
        \left(-\frac{t^2}{2\sigma_{\rm p}^2}\right)}
        \sin{\left(\omega_{\rm p}t\right)} \,
        \mathbf{e}_{\rm x} \,.
  \end{equation}
Here, $E_0$ is the pump amplitude, $\sigma_{\rm p}$ the pump
width, $\omega_{\rm p}$ the pump frequency, and
$\mathbf{e}_{\rm x}$ the unit vector along the chain.

\section*{Results}
Insight on how the lattice is set in motion can be obtained from
an effective phonon potential. By integrating the force in
Eq.~(\ref{eq:force_scaled}), we obtain a potential for a single
lattice displacement,
\begin{equation}
  \label{eq:pot}
  V_i(\Delta_i)= \frac{1}{8\lambda\gamma}
                       \Delta_i^2 + \Delta_i(n_i(t) - 1/2) \, ,
\end{equation}
which is just the sum of the local phonon potential energy and
the electron-phonon energy. For the perfectly-dimerized chain,
there are two possible ground states which differ from each
other by the sign of $\Delta_i$ and $n_i-1/2$ at every site (see
the left and right inset in Fig.~\ref{fig:system}(b)). When
$n_i$ changes, it rotates the potential parabola around
$\Delta_i = 0$ because its steepness is given by the first
derivative at that point, i.e., $n_i(t) - 1/2$. The equilibrium
condition of the zero force at the potential minimum gives
$\Delta_i^{\rm eq} = -4\lambda\gamma(n_i^{\rm
eq}-1/2)$ as well as the initial energy barrier
$\Delta V = 2\lambda\gamma(n_i^{\rm eq} - 1/2)^2$
which separates the two ground states in
Fig.~\ref{fig:system}(b). The equilibrium is perturbed by the
pump, which modifies $n_i^{\rm eq}$ and sets a new dynamical
minimum
$\Delta ^{\rm min}_i(t) = -4\lambda\gamma(n_i(t) - 1/2)$
(see the green dots in Fig.~\ref{fig:system}(b) and in the two
SI videos) towards which $\Delta_i(t)$ starts to move.
The pump is also modifying the initial energy barrier, making
$\Delta V(t)$ time-dependent. The barrier can completely
disappear for $n_i(t) = 1/2$, allowing the lattice to transition
from one excited ground state to another by flipping the sign of
$\Delta_i$.

\begin{figure}[tb]
\begin{center}
\includegraphics[width=0.49\textwidth]{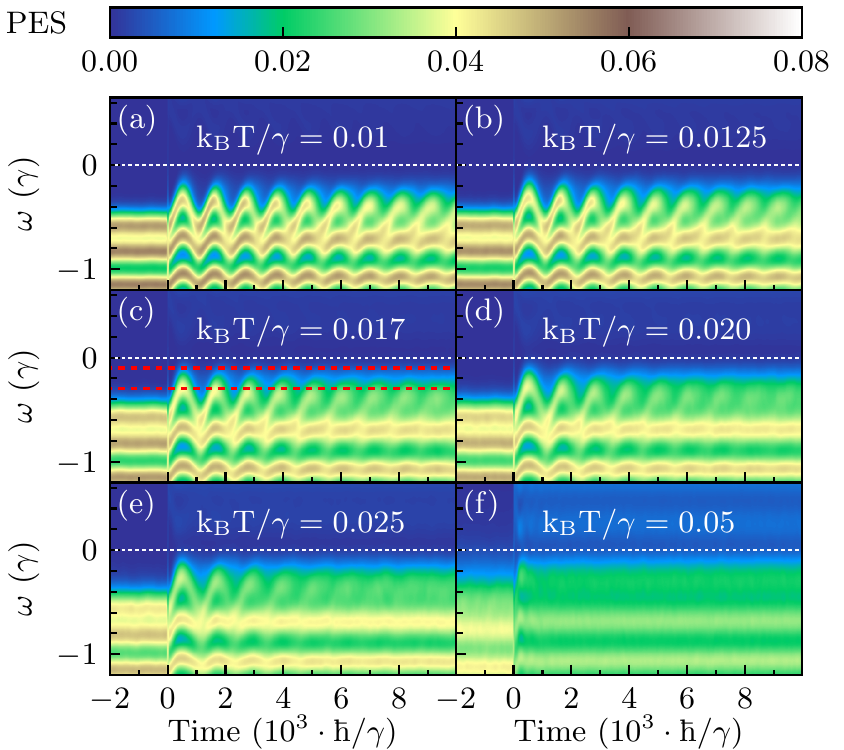}
\end{center}
\caption{\label{fig:PES}
  Photoemission spectrum for a chain of $L = 30$ sites at
  different temperatures. The electron-phonon coupling is
  $\lambda = 0.6$ and the phonon frequency is $\Omega = 0.01\
  \gamma / \hbar$. The pump parameters are $E_0 = 0.33$,
  $\sigma_{\rm p} = 10\ \hbar/\gamma$, and $\omega_{p} = 0.1\
  \gamma/\hbar$. The thin red dashed lines in panel (c) show the
  energy range where we average the PES in order to compute the
  inverted intensity shown in Fig.~\ref{fig:order}(c).
}
\end{figure}

As observed in experiments (see, e.g.,
Ref.~\onlinecite{schmitt_sci2008}), the lattice motion causes
the gap energy to oscillate, which is reflected in the computed
PES at low temperatures in Fig.~\ref{fig:PES}. The equilibrium
spectrum before the pump is gapped due to the Peierls distortion
and only the lower band is populated. After excitation, the
system will stabilize to a new equilibrium with a reduced gap
energy. Increasing the temperature of the lattice has two major
effects on the computed {PES}. The first is the evident damping
of the gap oscillations, so the system relaxes faster to a new
equilibrium PES for higher temperatures. The second effect is
the "washing out" of the finer details in the spectrum at higher
temperatures, increasing at longer times. The period of initial
PES oscillations in Fig.~\ref{fig:PES}(a) is much larger (around
1000 $\hbar /\gamma$) than what one would expect for $\Omega =
0.01$ $\gamma / \hbar$ which is a clear sign that
electron-phonon interaction significantly modifies the intrinsic
phonon frequency. The advantage of our self-consistent MC
approach is evident from the time scale of Fig.~\ref{fig:PES},
where the time resolution must be kept at $0.1\ \hbar / \gamma$
to capture the electron dynamics, still the fast evolution
scheme allows us to average over 3000 MC configurations.
Translating these units to the ones in experiments, for $\gamma
= 1$~eV, the time step is $\Delta t \approx 0.07$~fs, while the
simulation time is around 7 ps. For comparison, the
time-dependent density-matrix renormalization group method can
only reach times on the order of several
femtoseconds.~\cite{stolpp_prb, hashimoto_prb, matsueda}

\begin{figure}[tb]
\begin{center}
\includegraphics[width=0.49\textwidth]{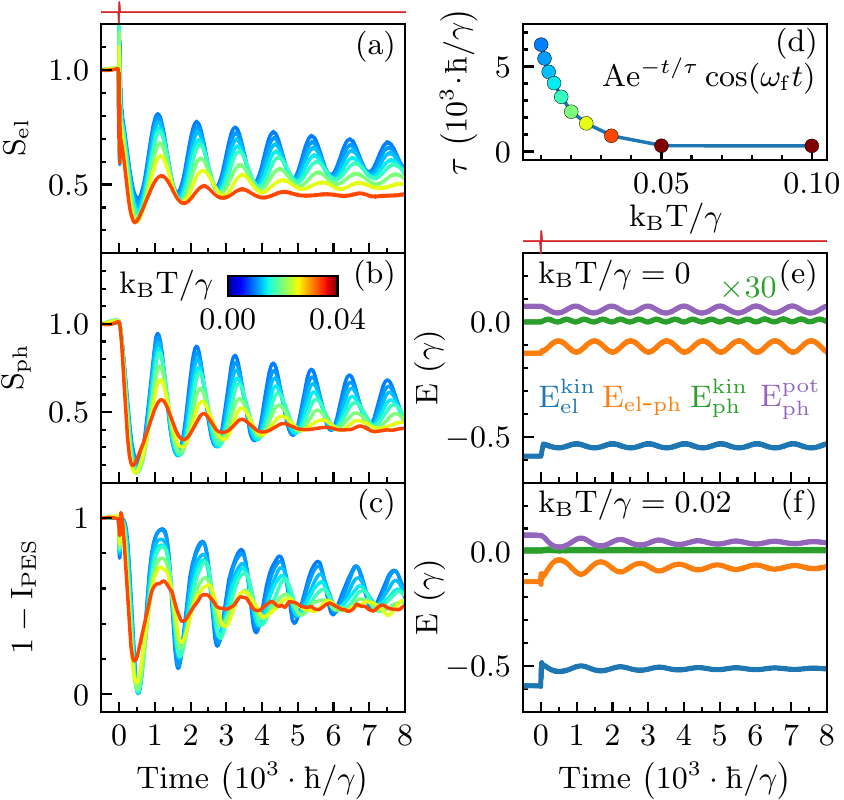}
\end{center}
\caption{\label{fig:order}
  (a), (b) Time evolution of the electron and phonon order
  parameters $S_{\rm el}(t)$ and $S_{\rm ph}(t)$, as defined in
  Eqs.~(\ref{eq:el_order}) and (\ref{eq:ph_order}), for different
  initial temperatures (see the colorbar in panel (b)). The
  order parameters are normalized to one before the pump is
  applied. (c) Time evolution of the inverse of a normalized
  in-gap PES intensity (averaged over the energy window between
  the two red lines in Fig.~\ref{fig:PES}(c)) for the same
  temperatures as in (a) and (b). (d) Temperature dependence of
  the relaxation time $\tau$ extracted from panel (a) by fitting
  $S_{\rm el}(t)$ to the form $Ae^{-t/\tau}\cos(\omega_{\rm f}
  t)$; the relaxation time diverges as $T\to 0$. (e), (f)
  Time evolution of the energy expectation values per site of
  the different contributions in Eqs.~(\ref{eq:ele}) and
  (\ref{eq:phe}),
  $E_{\rm el}^{\rm kin}$, $E_{\rm el-ph}$, 
  $E_{\rm ph}^{\rm kin}$, and $E_{\rm ph}^{\rm pot}$,
  at temperatures $k_{\rm B}T = 0$ and $k_{\rm B}T = 0.02\
  \gamma$. Due to the fast electron motion at $T=0$ (due to no
  relaxation), the electronic energies in panel (e) are time
  averaged over 50 time steps (with a sliding time window).
  These time-averaged curves are similar to the nonzero $T$
  curves in (f) except they do not get damped. The time profile
  of the pump field is shown as a thin red curve above the
  panels (a) and (e). The pump parameters are $\omega_{\rm p} =
  0.1\ \gamma/\hbar$, $\sigma_{\rm p} = 10\ \hbar/\gamma$, $E_0
  = 0.33$, while for the phonons $\Omega = 0.01\ \gamma/\hbar$
  and $\lambda = 0.6$.}
\end{figure}

The internal dynamics captured by the PES in Fig.~\ref{fig:PES}
is also reflected in two order parameters that track the
behaviour of the electronic and the lattice subsystem. For the
electrons, we define the order parameter via the time-dependent
density-density correlation function at the ordering vector
$q=\pi$, \begin{equation}
  S_{\rm el}(t) =
  \frac{1}{L} 
  \sum_{j_1, j_2}
  (-1)^{j_1 - j_2}
  \langle \hat{n}_{j_1}(t) \hat{n}_{j_2}(t)\rangle \,.
  \label{eq:el_order}
\end{equation}
For the lattice, we define a similar correlation function of the
local lattice displacements,
\begin{equation}
  S_{\rm ph}(t) =
  \frac{1}{L} 
  \sum_{j_1, j_2}
  (-1)^{j_1 - j_2}
  \langle \Delta_{j_1}(t) \Delta_{j_2}(t)\rangle \,.
  \label{eq:ph_order}
\end{equation}
The oscillations of these two order parameters, shown in
Figs.~\ref{fig:order}(a) and~\ref{fig:order}(b) for different
temperatures, follow the average inverted PES intensity near the
gap in Fig.~\ref{fig:order}(c). As with the PES in
Fig.~\ref{fig:PES}, the order parameters are also stabilizing at
a new equilibrium, with relaxation times inversely depending on
the initial lattice temperature. For each initial temperature in
Fig.~\ref{fig:order}(a), we fit the time dependence of $S_{\rm
el}(t)$ with an exponentially decaying function $A \,
e^{-t/\tau}\cos(\omega_{\rm f} t)$ to determine the relevant
relaxation time $\tau$. The fitted decay times in
Fig.~\ref{fig:order}(d) show that an initially hotter system
relaxes faster towards the new equilibrium.

We observe similar relaxation dynamics for the electron and
phonon energies in Fig.~\ref{fig:order}(f). During relaxation,
the electrons exchange kinetic energy with the lattice thus
reducing the initial lattice displacement and thereby also the
phonon potential energy $E^{\rm pot}_{\rm ph}$. The resulting
oscillations in Fig.~\ref{fig:order}(f) are underdamped. The
transfer of energy from electrons to phonons is a slow process,
happening on a time scale significantly longer than those
determined by relevant electronic and phonon energy scales. Due
to the self-consistent nature of our method we are able to
predict these nontrivial time scales starting only from a
tight-binding Hamiltonian in Eqs.~(\ref{eq:ele})
and~(\ref{eq:phe}) and without any additional assumptions about
the given system. Note that the oscillating behaviour can
already be deduced from the zero-temperature results with a
sliding time average, as shown in Fig.~\ref{fig:order}(e), but
the dynamics remains undamped and cannot give any insight into
relaxation times. Because of this, we apply our method at $T =
0$ only to describe the dynamics during or right after the pump
excitation, when damping effects are small, and we additionally
smooth some of the electronic observables by computing their
time averages. Further information on the raw results at $T = 0$
are provided in the SI.

\begin{figure}[bt]
\begin{center}
\includegraphics[width=0.49\textwidth]{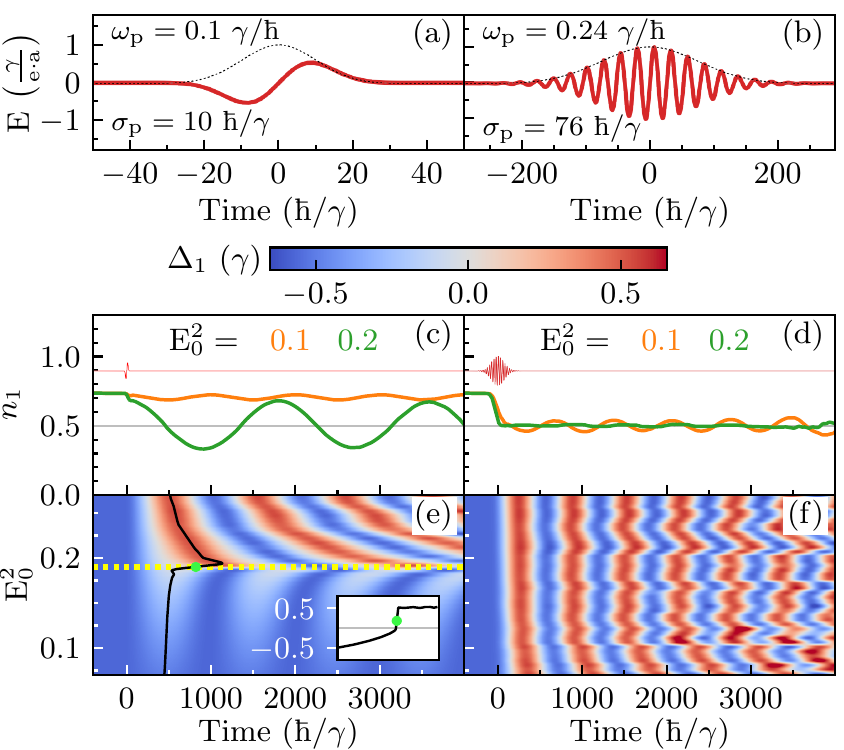}
\end{center}
\caption{\label{fig:melt}
  Perturbation of CDW order for a system starting at
  $k_\mathrm{B} T = 0$ by a pump. (a), (b) Electric field 
  (red curve) and its envelope function (dotted black curve)
  for two considered pump profiles: (a) the pump used in 
  Figs.~\ref{fig:PES} and~\ref{fig:order}, and (b) a pump
  with parameters set from Ref.~\onlinecite{schmitt_sci2008}.
  The field is given in units of $\gamma/(e\cdot a)$, where
  $e$ is the electron charge and $a$ is the lattice constant.
  (c), (d) Time evolution of the charge density $n_1(t)$ on the
  first site of the chain for the two pump pulses in (a) and
  (b) and for two different amplitudes. The thin red lines
  illustrate the time profiles of the two pulses from
  (a) and (b). The presented charge density is time-averaged
  over a window of 50 time steps. We refer the reader to the SI
  for raw $T = 0$ results. (e), (f) The time evolution of the
  phonon displacement $\Delta_1(t)$ on the first site of the
  lattice as a function of the pump intensity $E_0^2$ for
  the two pump profiles in (a) and (b), respectively.
  The dashed yellow line in (e) is the threshold intensity at
  which the lattice order changes sign. The inset in (e) tracks
  the amplitude of the first peak in $\Delta_1$ as marked with
  the black curve in (e). The phonon frequency is 
  $\Omega = 0.01 \gamma / \hbar$, $\lambda = 0.6$, and 
  $L = 600$.
}
\end{figure}

We examine the conditions for the weak and strong driving
scenarios by modulating the square of the pump amplitude $E_0^2$
(which is proportional to its fluence) for the two pulse
profiles in Figs.~\ref{fig:melt}(a) and~\ref{fig:melt}(b). For
the first profile, weak (strong) driving is illustrated in
Fig.~\ref{fig:melt}(c) for $E_0^2 = 0.1$ ($E_0^2 = 0.2$). The
inversion of order at $E_0^2 = 0.2$ happens in both electronic
and lattice subsystems simultaneously. To invert the order, the
pump needs to exceed a threshold intensity (marked by the dashed
yellow line in Fig.~\ref{fig:melt}(e)). 
The pump does not invert the order immediately, but it is the
coupled dynamics of electrons and phonons which leads to the
inversion
only at later times (e.g.~around 500 $\hbar/\gamma$ for
$E^2_0=0.2$ in Fig.~\ref{fig:melt}(c)), because the lattice
moves much slower than the electrons. At the crossover
intensity, the phonon amplitude saturates to a constant value
(see the inset in Fig.~\ref{fig:melt}(e)) and does not further
change with $E_0^2$. This amplitude saturation as a function of
fluence is one of the signatures of order inversion to look for
in CDWs but also in other systems with degenerate ground states
such as excitonic insulators, as recently demonstrated in
Ref.~\onlinecite{ning_prl_20}. 

We contrast the two driving regimes
obtained from the pump in Fig.~\ref{fig:melt}(a) with another
pump profile illustrated in Fig.~\ref{fig:melt}(b) which has
been used to study CDW melting in
Ref.~\onlinecite{schmitt_sci2008}. In the melting regime, the
electrons are almost instantly driven to a state with charge
density oscillating around 1/2 (Fig.~\ref{fig:melt}(d)). The
electron-phonon coupling at this point is effectively zero and
the electrons and phonons are dynamically decoupled for a period
of time. Therefore, phonons will initially oscillate fully
harmonically, as a sine wave with periodicity $2\pi/\Omega$ (see
Fig.~\ref{fig:melt}(f)), just to be perturbed by the electrons
at later times. The dependence on the pump intensity in
Figs.~\ref{fig:melt}(e) and~\ref{fig:melt}(f) was computed for
$T = 0$ because of numerical efficiency. However, the results
for low temperatures would be similar because the damping effect
caused by the temperature is low in the time range considered in
Figs.~\ref{fig:melt}(e) and~\ref{fig:melt}(f). Similar to the
electronic energy in Fig.~\ref{fig:order}(e), the charge density
at $T = 0$ is time-averaged; for raw data we refer the reader to
the SI.

\section*{Discussion}

We demonstrated that our time-dependent semi-classical MC
modelling captures the experimentally observed dynamics of
electron-phonon coupled systems driven out of equilibrium,
starting from the initial excitation to the subsequent
relaxation at finite temperatures. We have concentrated on an
indirect driving mechanism for the lattice dynamics. In future
studies, a direct coupling of the lattice to the field may give
additional insight into the interplay between indirect and
direct driving mechanisms. Moreover, different types of
electron-phonon interaction can lead to a periodic lattice
distortion accompanied by charge order, which opens up new
questions about their pump-probe dynamics. For example, in the
Su-Schrieffer-Heeger model~\cite{ssh_model} the diagonal
coupling of lattice displacements and onsite charge density is
replaced with an off-diagonal coupling between deviations of
neighboring bond length and the neighboring hopping. However,
the equations of motion remain of the same form, therefore the
indirect driving mechanism revolves around pump driving a local
bond current, which modifies the local bond length. Another
question is how the indirect driving mechanism is influenced by
anharmonic potential terms like $\sum_{i}\Omega_{4}\Delta_i^4$
considered in Ref.~\onlinecite{paleari_prb_2021}. Although this
term will modify the shape of the potential energy curve for
each lattice displacement, the connection between the
time-dependent charge density $n_i(t)$ and the local
displacement minima still exists. In addition, this approach
can be directly extended to two dimensions,~\cite{weber_2021}
but with additional computational cost.

Our explanation of the indirect driving in terms of the
modification of the local lattice potential and the reduction of
the energy barrier is very similar to the standard
phenomenological Ginzburg-Landau (GL) model, often employed to
explain trARPES experiments. The GL model predicts a double-well
structure in the free-energy potential, which upon excitation
with a strong pump can turn into a single-well potential where
the order parameter oscillates around
zero.~\cite{maklar_ncomm2021} Here, we propose an efficient
complementary approach to GL, which allows for a more detailed
microscopic exploration with consideration of the full
Hamiltonian of the system, but without the computational cost of
a more demanding first-principles method such as for example
time-dependent density functional theory. We view our method as
a balanced alternative which offers a way to self-consistently
induce lattice motion by direct coupling with the pump field and
to evolve the system to timescales relevant for phonon effects
to appear. The damping of the phonon oscillations emerges
naturally from our finite-temperature method, without the need
to modify the lattice equations of motion in order to include
damping. Another important characteristic of our time evolution
scheme is that it produces a true nonequilibrium electronic
state, without any additional assumptions regarding the time
scales it takes for electrons to thermalize. This opens up the
possibility to examine many different types of pump-probe
experiments.

The inversion of the lattice order, accompanied by the inversion
of electronic charge density ordering, is not a specific feature
of just CDW systems. A similar mechanism was recently reported
and measured in excitonic insulators.~\cite{ning_prl_20} A
double pump pulse was used to modulate the phonon oscillations
and suppress or enhance the phonon amplitude, and the
enhancement was related to the inverted structural order. A
similar modulation of the phonon amplitude was also reported for
tritellurides,~\cite{rettig_faraday2014} although it was not
directly associated with the inversion of order. The possibility
to invert the order appears to be a general feature of all
systems with degenerate ground states coupled with the lattice.
An interesting feature of the order inversion is that a system
can briefly pass through a metastable state where electrons and
phonons are decoupled, so the inversion of order is accompanied
with a state that might have dramatically different
conductivity.

\section*{Materials and Methods}
At $T=0$, the perfectly-dimerized state with
$\Delta_{0,i}=(-1)^i \Delta$ remains a static solution of the
equations of motion if no field is applied, whereas our MC
averaged quantities at $T>0$ require a thermalization period
before we apply the pump. For all our simulations, we use a time
step of $\Delta t = 0.1\ \hbar/\gamma$ which is set by the fast
electron dynamics.

The pump field modifies the phase $\varphi(t)$ of the
nearest-neighbor hopping in Eq.~(\ref{eq:ele}) through the
time-dependent vector potential
$
   \mathbf{A}(\mathbf{r}, t) =
   -c\int^t \mathbf{E}(\mathbf{r}, t') \, dt'
 $;
a spatially homogeneous field essentially shifts the electron
momentum. We work in the gauge where the time-dependent scalar
potential $\Phi(\mathbf{r}, t)$ is zero. 

The time-dependent PES is computed according to
Ref.~\onlinecite{freericks_pes}, 
\begin{align}
  P(\omega, t)
     = & -i \int_{-\infty}^{\infty}
   dt_1 \int_{\infty}^{\infty}dt_2\, s(t_1-t) s(t_2-t)
   e^{-i\omega(t_1-t_2)} \nonumber \\
    & \times \frac{1}{L}
   \sum_{i=1}^{L} G^{<}_{ii}(t_1, t_2) \, ,
\end{align}
where 
$G^{<}_{ii}(t_1, t_2) =
     i\langle 
     \hat{c}_{i}^{\dagger}(t_2)\hat{c}_{i}(t_1)
     \rangle$
is the time-displaced local lesser Green's function and
$
s(t) =
         \exp\left[ -{t^2}/
                         {2{\sigma}^{2}_{\rm probe}}\right]
               / \sqrt{2\pi}\sigma_{\rm probe}
                         $
is the shape of the probe pulse of width $\sigma_{\rm probe} =
10\ {\rm \hbar/\gamma}$ centered at $t=0$.

\section*{ACKNOWLEDGEMENTS}
  This work was supported by the U.S. Department of Energy
  (DOE), Office of Science, Basic Energy Sciences (BES) under
  Award DE-FG02-08ER46542. J.K.F.~was also supported by the
  McDevitt bequest at Georgetown University. This research used
  resources of the National Energy Research Scientific Computing
  Center (NERSC), a U.S. Department of Energy Office of Science
  User Facility operated under Contract no.~DE-AC02-05CH11231.


%

\setcounter{equation}{0}
\setcounter{figure}{0}
\setcounter{table}{0}
\makeatletter
\renewcommand{\theequation}{S\arabic{equation}}
\renewcommand{\thefigure}{S\arabic{figure}}

\onecolumngrid
\pagebreak
\setcounter{page}{1}
\section*{%
Supporting Information Appendix:
Theoretical description of time-resolved photoemission in
charge-density-wave materials out to long times}

\subsection*{Quantum versus classical harmonic oscillator}

The quantum and classical harmonic oscillators are closely
related to each other, especially if the harmonic oscillator has
a large average energy. We state some of the facts about the
quantum and classical oscillator that show this relationship. 

First, if the quantum state is described via a coherent state,
then the time dependence of the average position and the
momentum of the quantum oscillator are identical to that of a
classical pulled mass on a spring. Second, if we compute the
product of the fluctuations about the mean of the position and
momentum over a period of oscillation for a classical harmonic
oscillator it is given by $E/\Omega$, which is exactly what the
quantum state uncertainty product is for energy
eigenstates---the major difference is that the quantum
oscillator only has an allowed set of energies. Third, if the
average energy of the oscillator is larger than the phonon
frequency, then the difference between the quantum and classical
expectation values become quite small and decrease as the
average energy increases. The main difference is that a
classical oscillator can have energies smaller than
$\frac{1}{2}\hbar\Omega$, and the fluctuations in position and
momentum then both go to zero. This is clearly quite different
from that of the quantum oscillator. Hence, in situations where
the oscillator has a relatively large average energy, as
compared to $\hbar\Omega$, the semiclassical approximation for
the electron-phonon coupled problem should be quite accurate.
This should always occur when one pumps significant energy into
a system to drive it to nonequilibrium, as we do in the work
presented here.

\subsection*{Photoemission at zero temperature}

\begin{figure}[htb]
\begin{center}
\includegraphics[width=0.49\textwidth]{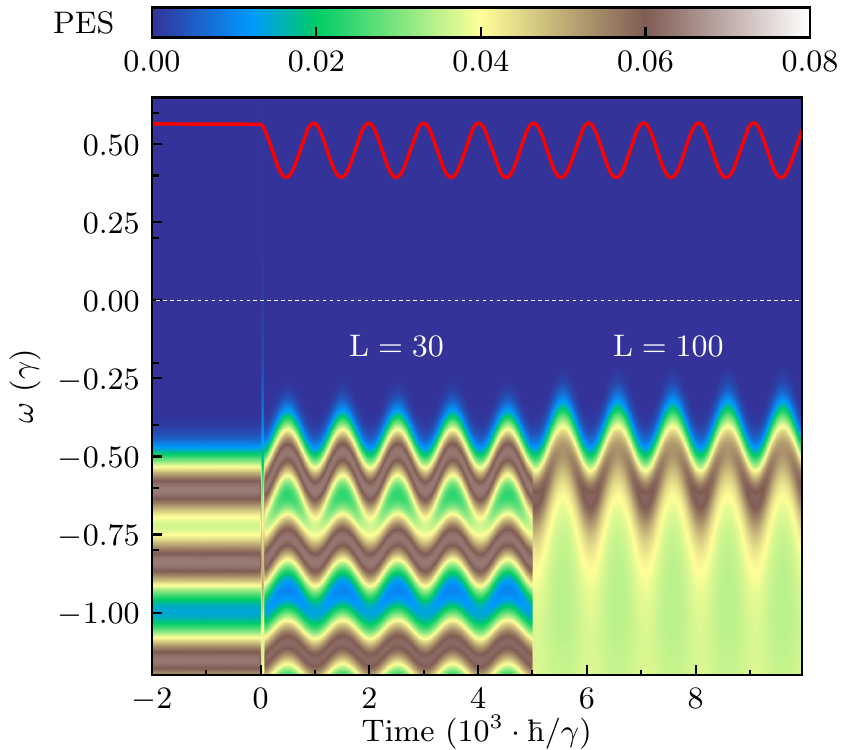}
\end{center}
\caption{\label{fig:S1}
  The photoemission spectrum at $T = 0$ temperature. Data for
  times below $5\cdot 10^3 \rm\ \hbar/\gamma$ is obtained for a
  system with $L = 30$ sites, while data above this time is
  obtained using a system with $L = 100$ sites. The red curve
  shows the oscillations of the phonon displacement, which
  corresponds to the gap energy $\Delta(t)$. Other parameters
  are the same as those used in Fig.~{2} in the main text:
  $\Omega = 0.01\ \gamma/\hbar$, $\lambda = 0.6$,
  $E_{0} = 0.33$, $\sigma_{\rm p} = 10\ \hbar/\gamma$,
  $\omega_{\rm p} = 0.1\ \gamma/\hbar$.}
\end{figure}

As already stated in the main text, results for zero temperature
are limited because they do not show any damping, and once the
system is excited, it will continue to oscillate indefinitely.
The photoemission spectrum at $T = 0$ is presented in
Fig.~\ref{fig:S1} The dynamics of energy levels reveals the gap
centered at $\omega= 0$ which is modulated by the symmetric
motion of the upper and the lower band, a behaviour previously
observed experimentally.~\cite{rettig_faraday2014} Note that
the discreteness of the spectra in Fig.~\ref{fig:S1} is due to
limited lattice sizes. We verify that $L=30$ sites are
sufficient to reliably estimate the oscillating gap and that the
PES becomes continuous for $L = 100$, as shown on the right of
Fig.~\ref{fig:S1}.

\begin{figure}[t]
\begin{center}
\includegraphics[width=0.70\textwidth]{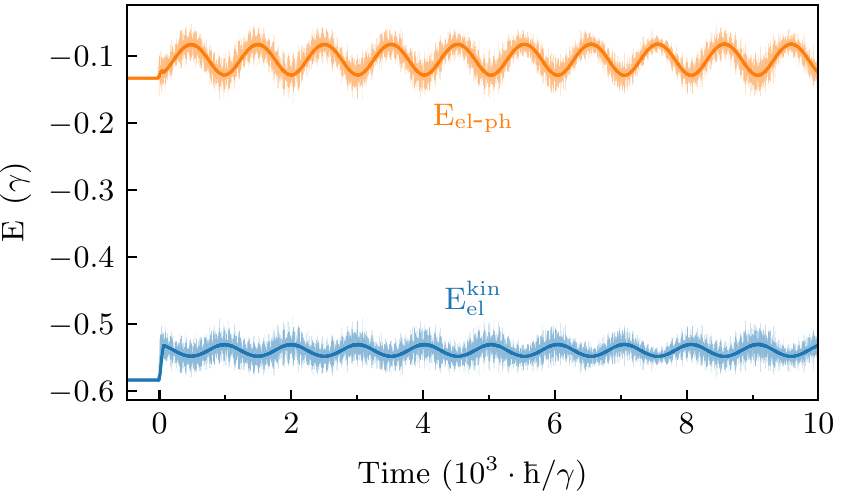}
\end{center}
\caption{\label{fig:S2}
 Raw $T = 0$ results for the electron kinetic and potential
 energy (transparent background) and the corresponding
 time-averaged results (full color curves) originally presented
 in Fig.~{3}(e) in the main text. The system parameters are: 
 $L = 30$, $\lambda = 0.6$, $\Omega = 0.01\ \gamma / \hbar$, 
 $E_0 = 0.33$, $\omega_{\rm p} = 0.1\ \gamma/\hbar$,
 and $\sigma_{\rm p} = 10\ \hbar / \gamma$.
 }
\end{figure}

\begin{figure}[bh]
\begin{center}
\includegraphics[width=0.70\textwidth]{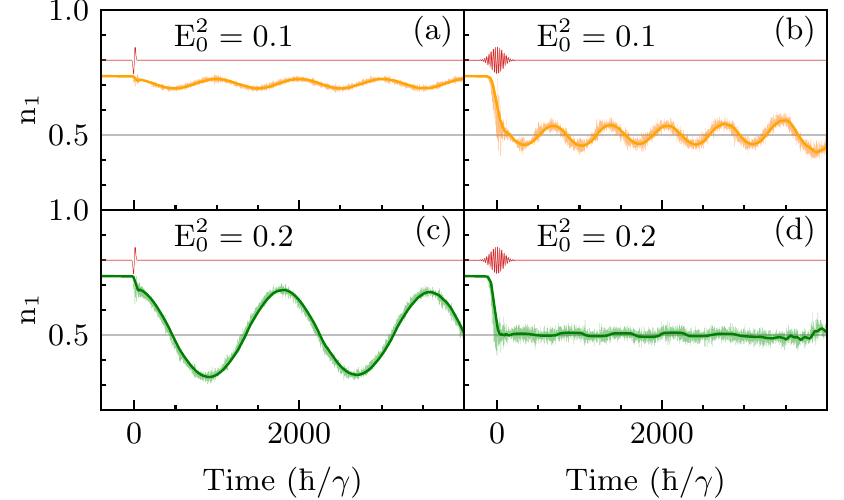}
\end{center}
\caption{\label{fig:S3}
  Weak and strong pump driving at $T = 0$ temperature.
  Full color curves in (a), (c) and (b), (d) correspond to
  time-averaged oscillations of the charge density at the 
  first site of the system, originally presented in Figs.~4(c)
  and~4(d) in the main text (for pump amplitudes $E^2_0 = 0.1$
  and $E^2_0 = 0.2$), respectively. The transparent background
  shows the raw data before the time averaging. Other system
  parameters are: $L = 600$, $\lambda = 0.6$, $\Omega = 0.01
  \ \gamma / \hbar$. The system in (a) and (c) is excited using
  a pump with $\omega_{\rm p} = 0.1\ \gamma/\hbar$,
  $\sigma_{\rm p} = 10\ \hbar / \gamma$, while that in
  (b), and (d) is excited with a pump with
  $\omega_{\rm p} = 0.24\ \gamma/\hbar$,
  $\sigma_{\rm p} = 76\ \hbar / \gamma$.
}
\end{figure}

The raw data for the electronic energies at $T = 0$ presented in
Fig.~{3} in the main text is shown in Fig.~\ref{fig:S2}, while
the raw data for the charge density excitation for the weak and
strong pump shown in Figs.~{4}(c) and {4}(d) in the main text
are shown in the corresponding
Figs.~\ref{fig:S3}(a),~\ref{fig:S3}(c),
and~\ref{fig:S3}(b),~\ref{fig:S3}(d) respectively.
%
%

\begin{figure}
\centering
\includegraphics[width=0.75\textwidth]{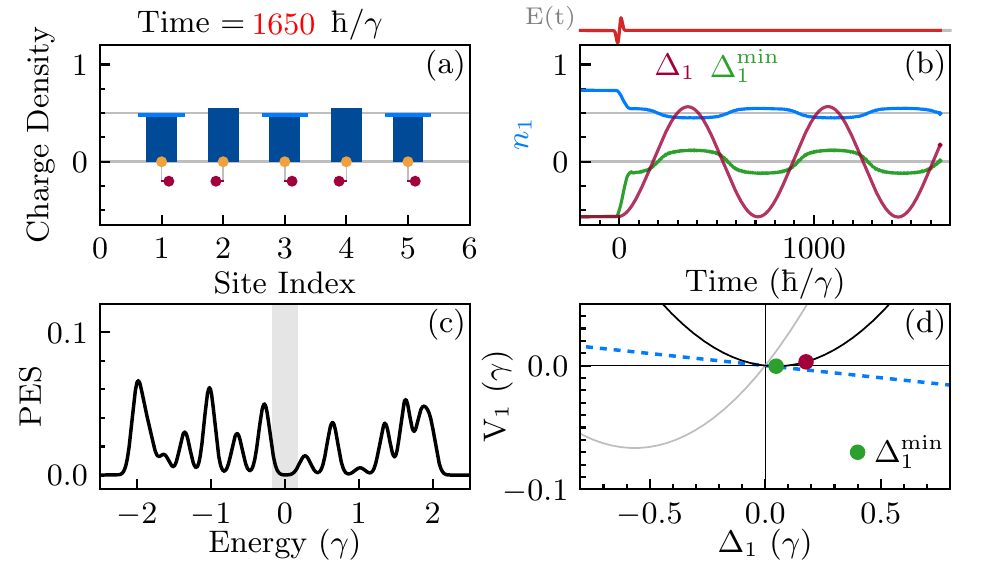}
\caption{%
  \label{fig:hdyn}
  Explanation of the data presented in the two
  supplementary movies
 (\href{https://figshare.com/s/4d6f903422282662e3ed}{Video 1}
  and
  \href{https://figshare.com/s/3c972f9386c9c1b82692}{Video 2}):
  (a) Local charge density $n_i(t)$ (dark blue bars) on the
  first five sites (orange dots). The lattice displacements
  $\Delta_i(t)$ (dark red dots and dark lines below the orange
  dots) are scaled from their original values to values below
  one half in order to fit them between the neighboring sites.
  Light blue horizontal lines above the blue bars emphasize the
  charge density $n_{i}(t)$ on odd sites. (b) Time-dependence of
  the phonon displacement $\Delta_1(t)$ (the dark-red curve),
  the minimal displacement determined from the potential energy
  $\Delta_1^{\min}(t)$ (the green curve) and the time-averaged
  charge density $n_1(t)$ (the light blue curve) on the first
  site (and all odd sites). (c) The intensity of the
  photoemission spectrum. The gray rectangle at the center shows
  the expected gap size obtained from $\Delta_1(t)$. (d) The
  potential energy profile for the first site
  $V_1(\Delta_1(t))$. The dark-red dot shows the actual
  position $\Delta_1(t)$ while the green dot show the
  $\Delta_1^{\min}(t)$. The dashed light blue line shows the
  tangent of the energy parabola determined from the first
  derivative $n_1(t) - 1/2$, the gray parabola in the background
  shows the energy profile in equilibrium before the pump. The
  electron charge density is time-averaged over 50 time steps.}
\end{figure}

\begin{figure}[hb]
\centering
\includegraphics[width=0.75\textwidth]{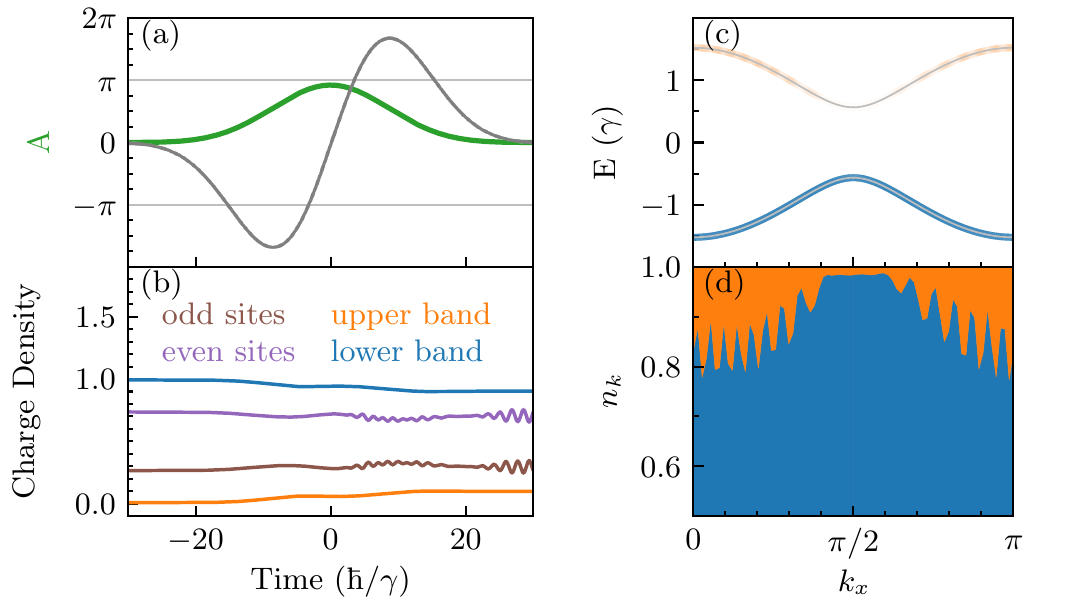}
\caption{%
  \label{fig:kspace}
  Explanation of the data presented in the 
  \href{https://figshare.com/s/b568afdf1f3355a6f18e}{Video 3}: 
  Time evolution of the electron
  occupation numbers in the upper and lower bands. (a) The time
  profile of the vector potential intensity ${A}(t)$ (the green
  curve) and the pump field $E(t)$ (the gray curve). (b) The
  time profile of the average charge density on all even sites
  (the violet curve) and all the odd sites (the brown curve) in
  real space. The corresponding Fourier transformed values show
  the time profiles of the electron occupation numbers in the
  lower band (the blue curve) and in the upper band (the orange
  curve). (c) $k$-resolved occupation of the upper and lower
  bands. The transparency of the blue and orange curves
  corresponds to the occupation number for a given $k$ (full
  transparent equals to zero occupation). (d) Graphical
  representation of the $k$-resolved occupation shown in (c).
  The orange surface corresponds to the upper band, while the
  blue surface corresponds to the lower band. Since the vector
  potential shifts the electron momentum, we translate the
  results in (c) and (d) with $\mathbf{A}(t)$ to keep them gauge
  invariant.}
\end{figure}

Figures~\ref{fig:hdyn} and~\ref{fig:kspace} additionally explain
the data presented in the three supplementary videos.

{\label{mov:weak} 
\href{https://figshare.com/s/4d6f903422282662e3ed}{Video 1}:
 Time evolution of the electronic and the lattice system
for the weak pump at $k_{\rm B}T = 0$ temperature. The presented
data is obtained for $T = 0$ temperature because of numerical
efficiency. The fast oscillations of the charge density are
expected to average to smooth curves for $T > 0$ which would be
closer to the presented time-averaged $n_1(t)$. We have
verified this by doing calculations at a finite temperature as
seen by comparing Fig.~3(e) and Fig.~3(f) in the main text.
Parameters used are $E_0 = 0.33$, $\lambda = 0.6$, $\sigma_{\rm
p} = 10\ \hbar/\gamma$, $\omega_{\rm p} = 0.1\ \gamma/\hbar$, 
$\Omega = 0.01\ \gamma/\hbar$, $\gamma = 1$, $L = 30$.}

{\label{mov:strong} 
\href{https://figshare.com/s/3c972f9386c9c1b82692}{Video 2}:
Time evolution of the electronic and the lattice system for
the strong pump at $k_{\rm B}T = 0$ temperature. The presented
data is obtained for $T = 0$ temperature because of numerical
efficiency. The fast oscillations of the charge density are
expected to average to smooth changes for $T > 0$ which would
be closer to the presented time-averaged $n_1(t)$.
Parameters used are $E_0 = 0.66$, $\lambda = 0.6$,
$\sigma_{\rm p} = 10\ \hbar/\gamma$, $\omega_{\rm p} = 0.1
\ \gamma/\hbar$, $\Omega = 0.01\ \gamma/\hbar$,
$\gamma = 1$, $L = 30$.}

{\label{mov:ks}
\href{https://figshare.com/s/b568afdf1f3355a6f18e}{Video 3}: 
Electron dynamics in $k$-space during the pump
excitation at $k_{\rm B}T = 0$. Here,
$E_0 = 0.4$, $\Omega = 0$, $\lambda = 0.6$,
$\sigma_{\rm p} = 10\ \hbar/\gamma$,
$\omega_{\rm p} = 0.1\ \gamma/\hbar$, $\gamma = 1$.
}

\end{document}